\documentclass[pra,aps,showpacs,floatfix,twocolumn
]{revtex4-1}
\usepackage{graphicx}
\newcommand{\be}{\begin{eqnarray}}
\newcommand{\ee}{\end{eqnarray}}
\def\beq{\begin{equation}}
\def\eeq{\end{equation}}

\usepackage{amsmath}
\usepackage{amssymb}

\begin{document}
\title{Testing the universality of the many-body metal-insulator transition by time evolution of a disordered one-dimensional ultracold fermionic gas}
\author{Masaki Tezuka}
\email{tezuka@scphys.kyoto-u.ac.jp}
\affiliation{Department of Physics, Kyoto University, Kitashirakawa, Sakyo-ku, Kyoto 606-8502, Japan}
\author{Antonio M. Garc\'{\i}a-Garc\'{\i}a}
\email{amg73@cam.ac.uk}
\affiliation{Cavendish Laboratory, University of Cambridge, JJ Thomson Avenue,
Cambridge, CB3 0HE, UK}
\begin{abstract}
It is now possible to study experimentally the combined effect of disorder and interactions in cold atom physics.
Motivated by these developments we investigate the dynamics around the metal-insulator transition (MIT) in a  
one-dimensional (1D) Fermi gas with short-range interactions in a quasiperiodic potential by the time-dependent density-matrix renormalization group (tDMRG) technique. By tuning disorder and interactions we study the MIT from the weakly to the strongly interacting limit. The MIT is not universal as time evolution, well described by a process of anomalous diffusion, depends qualitatively on the interaction strength. By using scaling ideas we relate the parameter that controls the diffusion process with the critical exponent that describes the divergence of the localization length. In the limit of strong interactions theoretical arguments suggest that the motion at the MIT tends to ballistic and critical exponents approach mean-field predictions.
 \end{abstract}
\maketitle
Studies of the interplay of interactions and disorder have flourished in recent years \cite{fallani,flach,aiz,wang,fishman}. Reasons for this renewed interest include cold atom experiments \cite{fallani}, more quantitative numerical simulations \cite{flach} and novel theoretical techniques \cite{aiz,wang}. Adding further appeal to this problem, numerical results for interacting 1D bosons \cite{flach} in a disordered potential contradicts rigorous mathematical predictions \cite{wang}. 
Reasons for these discrepancies are not yet well understood \cite{fishman}.
 
Here we address a related problem: the time evolution of 1D fermions with short-range attractive interactions by tDMRG techniques.
We choose tDMRG over other techniques because the range of sizes that can be accessed is much larger.
This is key to minimize finite size effects that might obscure the occurrence of localization. We focus on dynamical properties as the time dependence of the atom distribution is a natural observable in cold atom experiments.
 Disorder is modelled by a quasiperiodic potential \cite{aubry} that can be implemented experimentally \cite{fallani}, 
\be
V(n) = \lambda \cos(2\pi \omega n +\theta)
\label{qp}
\ee
with $\omega$ irrational, $\theta \in [0,2\pi)$, and $\lambda > 0$. 
In the non-interacting limit a 1D tight-binding model with this potential and a hopping parameter $J\equiv 1$ 
 undergoes a MIT at $\lambda_\mathrm{c} = 2$ \cite{komo}. As attractive interactions are turned on $\lambda_\mathrm{c}$ decreases  \cite{uspra}. It is thus possible to study the role of interactions at the MIT from the weak to the strong coupling limit.
We employ the term MIT instead of superconducting-insulator transition because according to \cite{uspra} quasi-long-range order is already broken when the insulator transition occurs.

\noindent {\it The model.}---
We employ tDMRG \cite{White-Feiguin-2004} to study the dynamics of the $L$-site spin-$1/2$ Hubbard model,
\begin{eqnarray}
\mathcal H &=& -J\sum_{i=1,\sigma}^{L-1}(\hat c_{i-1,\sigma}^\dag
\hat c_{i,\sigma}+\mathrm{h.c.})
+ U \sum_{i=0}^{L-1} \hat n_{i,\uparrow}\hat n_{i,\downarrow}\nonumber\\
&+& \sum_{i=0}^{L-1} V(i) \hat n_{i},
\label{eqn:Hubbard}
\end{eqnarray}
in which
$\hat c_{i,\sigma}$ annihilates an atom at site $i$ in spin state $\sigma(=\uparrow, \downarrow)$,
$\hat n_{i,\sigma}\equiv \hat c_{i,\sigma}^\dag \hat c_{i,\sigma}$,
$\hat n_i\equiv \hat n_{i,\uparrow}+\hat n_{i,\downarrow}$
$U<0$ is the on-site interaction, and
$V(i)$ is given by (\ref{qp}) with $\omega = (\sqrt{5} - 1)/2$.
The angle $\theta$ is chosen so that $V(i)$ is symmetric relative to the center of the system.

The tDMRG  provides an efficient
way to simulate the time evolution of a wavefunction obtained with DMRG.
Our initial configuration ($t = 0$) is the ground state of the Hamiltonian where the disordered potential (\ref{qp}) is replaced by a simple 
potential well of width $\ell=64$ and depth $D=10$ centered at the origin, 
\be
V_{t<0}(i) = D\Theta(|x_i|-\ell/2),
\label{vtneg}
\ee
 where $\Theta$ is the Heaviside function and $x_i \equiv i - (L-1)/2$ is the location of the site relative to the center of the system.
For $t > 0$ we compute the real-time evolution ($t > 0$) of this ground state under the 
Hamiltonian $\mathcal{H}$ for $L=256$ after the potential well is replaced by the quasiperiodic potential (\ref{qp}).
$\mathcal{H}$ is broken into terms affecting only two neighbouring lattice sites.
The time evolution operator $e^{-i \mathcal{H} \Delta t}$, decomposed using the
second order Suzuki-Trotter breakup, is iteratively applied
on the ground state obtained by finite system DMRG.
The time step $\Delta t$, measured in units of $\hbar/J$, satisfies $0.01 \leq \Delta t \leq 0.05$ and $m = 200$ states have been
kept in the DMRG simulation unless noted otherwise.

Before we proceed with the calculation we provide a brief overview of previous research on this model.
In the non-interacting limit, $U = 0$, the MIT is described by a process of anomalous diffusion \cite{anodi} controlled by the multifractal dimensions of the spectrum \cite{komo}. The 
localization length $\xi \propto |\lambda - \lambda_\mathrm{c}|^{-\nu}$ diverges at the transition with $\nu \approx 1$ \cite{hashi}. 
 For $\lambda = 0$ the model is exactly solvable \cite{hubbard} for all $U$'s. For $|U| \gg 1$ it is mapped onto a weakly interacting hard-core Bose gas with a rescaled hopping parameter $J' \approx J^2/|U|$ \cite{shepe}. This suggests that the MIT will occur at $\lambda_\mathrm{c} \sim J^2/|U|$.
 
For finite disorder and interactions there are already several studies about the static properties of Eq. (\ref{eqn:Hubbard}) \cite{hiro,shiro,uspra} and related models \cite{magnu,modu,schreiber,chavez,giamarchi}. The dynamics of an interacting 1D Bose gas in a quasiperiodic potential was first investigated numerically in \cite{magnu}. For a more recent study in which interactions are treated in a mean-field fashion we refer to \cite{modu}. In \cite{uspra,shiro} it was found that, in a 1D Fermi gas with attractive interactions, $\lambda_\mathrm{c} = \lambda_\mathrm{c}(U)$ depends on the interaction and that weak disorder can enhance superfluidity.  Renormalization group techniques were employed in \cite{giamarchi} to study the weak disorder limit of spinless fermions in the Fibonacci chain, a quasiperiodic potential that is critical for each value of the coupling constant.
 For sufficiently weak interactions it was found in \cite{hiro} that the spectrum of the Fibonacci chain is still multifractal. However in the case of the potential (\ref{qp}) the system becomes an insulator for $U < 0$ and $\lambda = 2$.
 
\noindent {\it Results.}---
In order to investigate the dynamics of (\ref{eqn:Hubbard}) we first compute the $n$-th order moment
defined as
\be 
\langle x^n(t) \rangle \equiv \left[ \frac
{ \sum_i |x_i|^n \langle \Psi(t) | \hat n_i | \Psi(t) \rangle }
{ \sum_i \langle \Psi(t) | \hat n_i | \Psi(t) \rangle }
 \right],
 \label{x2t}
\ee
in which $|\Psi(t)\rangle$ is the many-body wavefunction at time $t$.
Here, $i=0,1,\ldots,L-1$ runs over the site index and
$\hat n_i = \sum_\sigma \hat c_{i,\sigma}^\dag \hat c_{i,\sigma}$ is the number
operator at site $i$.
We set $L=256$ and the number of fermions per spin to $N=12$.
Initially fermions are confined to sites $96$--$159$ by the potential well $V_{t<0}(i)$ (\ref{vtneg}).
 Then we study the time evolution after the potential well is removed and the quasipotential $V(i)$ is switched on at $t = 0$.

\begin{figure}[t]
\includegraphics{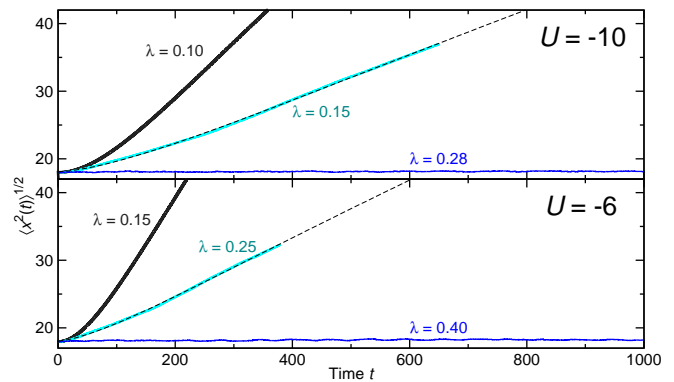}
\caption{(Color online)  $U=-10$: tDMRG calculation of $\langle x^2(t) \rangle$  Eq. (\ref{x2t}) for,  from top to bottom, $\lambda = 0.1, 0.15, 0.28$. Diffusion is clearly suppressed at $\lambda = 0.28$.  The dashed line 
is the best, $\alpha \approx 1.54$, fitting function (\ref{x2fit}) around the MIT $\lambda = \lambda_\mathrm{c} \approx 0.15$. 
  $U = -6$: tDMRG calculation of $\langle x^2(t) \rangle$ (\ref{x2t}) for,  from top to bottom, $\lambda = 0.15, 0.25, 0.40$.
The dashed line 
is the best, $\alpha \approx 1.48$, fitting function (\ref{x2fit}) around the MIT $\lambda = \lambda_\mathrm{c} \approx 0.25$.  In both figures $L=256$ and $N = 12$.
The maximum time $t_{\rm max}$ that we can explore is dictated either by the stability of the tDMRG simulation $(U,\lambda) = (-10,0.28),(-6,0.40)$ or by the growing importance of finite size effects for $t > t_{\rm max}$ in the rest of cases.    
   }
\label{x2}
\end{figure}

The results for $\langle x^2(t) \rangle$ and different $\lambda$'s are depicted in Fig.~\ref{x2}. 
The values $U=-6, -10$ correspond to the regime of strong coupling where the interaction energy is
 larger than the kinetic and potential energy due to the quasiperiodic potential. 
We clearly observe in Fig.~\ref{x2} arrest of diffusion for sufficiently large $\lambda$. The critical disorder $\lambda_\mathrm{c} < 2$ for which the MIT occurs decreases as $|U|$ increases. We have estimated $\lambda_\mathrm{c}$ directly from $\langle x^2(t) \rangle$ by identifying a narrow region of $\lambda$'s for which the dynamics becomes substantially slower than in the metallic region and also by an explicit calculation of the participation ratio \cite{uspra}. 
In the latter the critical $\lambda_\mathrm{c}$ at which MIT occurs, for a fixed $(L, U)$, is identified as a maximum of 
the participation ratio as a function of $\lambda$. We have also found that $\lambda_\mathrm{c}$ does not strongly depend on the filling factor provided that the chemical potential is far from the band edge.

\begin{figure}[pbt]
\includegraphics[width=0.82\columnwidth,clip,angle=0]{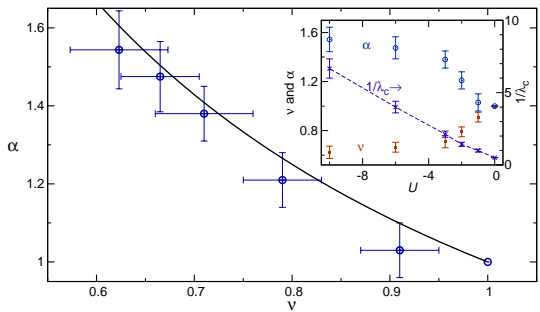}
\caption{(Color online) $\alpha$ (see text around (\ref{x2fit}) for details)  versus the critical exponent $\nu$. The latter is obtained by fitting (\ref{de})  to $\Delta E = a e^{bL|\lambda -\lambda_\mathrm{c}|^\nu}$ with $a,b,\nu$ fitting parameters, $L =13$ and $\omega = 5/13$ in  (\ref{qp}). Fitting is restricted to a small region $\lambda > \lambda_\mathrm{c}$ such that $\xi < L$. Error bars were obtained by considering the stability of the results under small changes in $\lambda_\mathrm{c}$ and the fitting interval.  
From left to right, the points correspond to $(U, \lambda_\mathrm{c}) = (-10,0.15),(-6,0.25),(-3,0.47),(-2,0.70),(-1,1.01),(0,2)$. Qualitative agreement with the expression $\nu = \frac{1}{\alpha}$ (solid line) is observed.
Inset: $\nu$ and $\alpha$ (left axis) and $1/\lambda_\mathrm{c}$ (right axis) plotted against $U$. 
The dashed line is a guide to the eye.}
\label{nup}
\end{figure}

In order to fit the numerical data we employ the ansatz,
\be
 \langle x^2(t) \rangle = x_0^2(1+(t/t_0)^\alpha)
 \label{x2fit}
 \ee 
 where $x_0$, $t_0$ and $\alpha$ are fitting parameters. We note that this fitting function is only an educated guess. We choose it because, despite its simplicity, it led to a good description of the data.   Other functions recently used in the literature \cite{fallani} were also tried but the fitting was qualitative worse. 
  Results of the best fit (see Fig.~\ref{x2}) are presented 
 in Fig.~\ref{nup} for different values of $U$ at $\lambda \approx \lambda_\mathrm{c}$. It is observed that $\alpha$ depends on $U$ and it is different from the one for $U=0$, $\alpha \approx 1 \approx 2 d_\mathrm{H}$ where $d_\mathrm{H}$ is the Hausdorff dimension of the spectrum \cite{komo}. Therefore strong interactions modify substantially the dynamics at the MIT.
 
This is an important result. According to the one parameter scaling theory \cite{gang4} the parameter $\alpha$  is related to the critical exponent $\nu$ that labels the universality class of the MIT. Therefore different $\alpha(U)$ at the MIT correspond to different universality classes.   
 An important concept of this theory is the dimensionless conductance $g = E_\mathrm{T}/\delta$ where $E_T$, the Thouless energy, is the energy related to the typical time for a particle to cross a sample of size $L$ and  $\delta$ is the mean level spacing. For a disordered metal (normal diffusion) $g(L) \propto L^{D-2} \to \infty$ for $L \to \infty$ since $E_T \propto 1/L^2$ and $\delta \propto 1/L^D$.  
Analogously for an insulator $g(L) \propto e^{-L/\xi}$ decays exponentially. A MIT is characterized by a scale independent $g(L) \equiv g_\mathrm{c}$. Two mechanisms can lead to this scale invariance: localization effects that slow down the motion $\langle x^2(t) \rangle \propto  t^{\alpha}$ at the MIT and a multifractal spectrum \cite{komo}, with the Hausdorff dimension $d_\mathrm{H}$, that induces an anomalous scaling of $\delta \propto 1/L^{D/d_\mathrm{H}}$. Based on these arguments it was predicted in \cite{gang4} that in $D = 1$ a MIT will occur provided that $2 d_\mathrm{H} = \alpha$. In the non-interacting limit this relation was verified in \cite{komo,anodi}.

In the presence of repulsive interactions it has been suggested \cite{macmillan} the scaling theory must include two parameters, $g$ and the ratio between an energy related to interactions and the mean level spacing. 
For attractive interactions, especially in a quasiperiodic potential, the situation is less clear. In any case the above arguments, together with the numerical results above, provide a rather compelling albeit qualitative picture of the role of interactions: as $|U|$ increases, $\alpha > 1$ increases and the motion becomes superdiffusive. According to the scaling theory, the MIT can occur only if $d_\mathrm{H}$ also increases. Physically that means that interactions smooth out the fractal properties of the spectrum at the MIT. The smoothing will be substantial when the interacting energy is much larger than the typical size of the subbands induced by the quasiperiodic potential around the Fermi energy. In this large $|U|$ limit, corresponding to hard-core bosons, the spectrum is no longer fractal ( $d_\mathrm{H} \approx 1$) and therefore the dynamics at the MIT $\alpha = 2d_\mathrm{H} \approx 2$ approaches the ballistic limit. The numerical findings of \cite{hiro} and the semi-analytical results of \cite{giamarchi} for spinless fermions fully support this picture. We note that \cite{aiz} many features of the many-body MIT are similar to those of a single particle in a Cayley tree \cite{cayley}.  For this model $\alpha = 2$ and $\nu = 1/2$ around the MIT. It is thus tempting to speculate that these results also applies to the Hamiltonian (\ref{eqn:Hubbard}) in the limit $|U| \to \infty$.

Before we turn to the next observable
a few comments are in order: a) the fitting interval is long enough for disorder and interactions to strongly influence the motion, b) the motion is slower as $|U|$ increases. The length of the fitting interval (see below) increases accordingly. 
As a result, for $|U| \gg 1$ the value of $\alpha$ is more dependent on the interval. It is thus likely that additional transient terms are present in (\ref{x2fit}). We stick to (\ref{x2fit}) because the addition of more terms without a clear physical motivation would lead to ambiguous results, 
c)  
the maximum time that we represent in the figures, and that it is used in the fittings, was chosen so that both the numerical error accumulation ($t \lesssim 1000$) and finite size effects that obscure localization are negligible. 
For the latter the maximum time strongly depends on $U, \lambda$. This maximum time $t_\mathrm{max}$ for which finite size effects are not important is chosen by imposing that the occupation number of the last five sites remains less than $0.01$ 
and no sharp increases occurs for smaller times. For instance, around $\lambda \approx \lambda_\mathrm{c}$,
$t_\mathrm{max} \approx 650$ for $U=-10$ but only
$t_\mathrm{max} \approx 75$ for $U=-1$,  d) the moments of the distribution, such as those depicted in Fig.~\ref{x2}, 
can be easily measured in cold atom experiments. Therefore this model is an ideal candidate for experimental tests of 
the MIT in strongly interacting 1D cold Fermi gases.

\begin{figure}[t]
\includegraphics{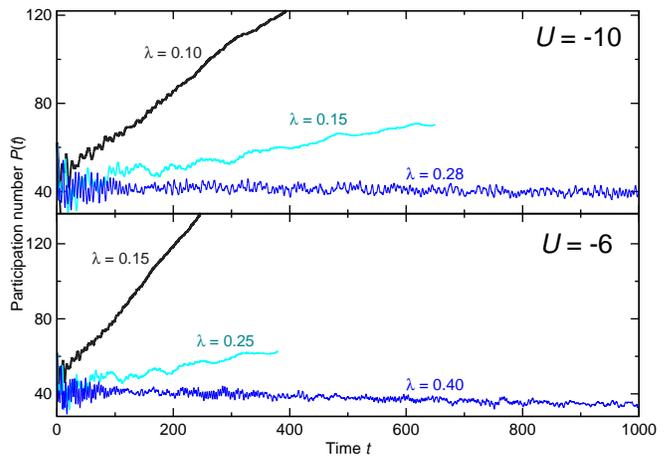}
\caption{(Color online) Participation number $P(t)$ (\ref{pt1}) for the same parameters as Fig. \ref{x2}. A non increasing $P(t)$ for $t \to \infty$ is a signature of localization.}
\label{part}
\end{figure}

In order to obtain information of the time evolution of the full many-body wavefunction 
we have also computed the time-dependent participation number
\cite{Kopidakis-Komineas-Flach-Aubry-2008},
\be
P(t) \equiv \frac
{ ( \sum_i \langle \Psi(t)|\hat n_i| \Psi(t) \rangle )^2 }
{ \sum_i \langle \Psi(t) |\hat n_i| \Psi(t) \rangle^2 }
\label{pt1}
\ee
which, up to normalization factors, gives an estimation of the number of sites which, at a given time,  are occupied (see \cite{Wegner1980} for more information). In an insulator $P(t)$ will be constant for sufficiently long times but in a metal it will always increase with time. Even a steady increase indicates that at least some parts of the wavepacket can escape localization.
Therefore $P(t)$ is an indicator of localization of the full wavepacket. In Fig.~\ref{part} we plot $P(t)$ for $U=-10, -6$ and different $\lambda$'s. The results are fully consistent with the previous calculation of moments. The transition is located around the same $\lambda_\mathrm{c}$ and no increase in time is observed for $\lambda \gg \lambda_\mathrm{c}$.

 We now turn 
 to further substantiate the non-universality of the MIT by an explicit calculation of the critical exponent $\nu$. For that purpose we study the sensitivity of the ground state to a change of boundary conditions \cite{schreiber}, 
 \be
 \Delta E = E_\mathrm{P} - E_\mathrm{A}
 \label{de}
 \ee 
 where $E_\mathrm{P}$ and $E_\mathrm{A}$ stand for the ground state for periodic and anti-periodic boundary conditions respectively. As the MIT is approached from the insulator side $\Delta E \propto e^{-L/\xi}$ with $\xi \propto |\lambda -\lambda_\mathrm{c}|^{-\nu}$. We exploit this relation to find $\nu$, with $L=13$. In Fig.~\ref{nup} we present results for $\nu(U)$ for different $\alpha(U)$ at $\lambda = \lambda_\mathrm{c}$. 
 It is observed that as $U$ increases $\nu$ decreases from its non-interacting value $\nu \approx 1$. This is an additional indication that the MIT in many-body systems is not universal. However the expected approach to the mean-field limit $\nu = 1/2$ for $U \gg 1$ seems to be slow. 
Theoretical arguments \cite{myprl,gang4} suggest that the anomalous diffusion, through $\alpha(U)$, at the MIT is directly related to the critical exponent $\nu(U)$ that labels the universality class of the MIT. The simplest expression consistent with ideas and techniques employed in the non-interacting limit \cite{hiro,hashi,myprl} is $\nu = \frac{1}{\alpha}$, which is in qualitative agreement (see Fig.~\ref{nup}) with the numerical results. Finally we note that the calculation of $\nu, \alpha$ is rather crude and subjected to substantial uncertainties in the fitting procedure. This is especially true for the $U=-6,-10$ for which the value of $\alpha$ is rather sensitive to both the fitting interval and the details of the fitting function (\ref{x2fit}). For instance for smaller intervals and fitting functions including additional transient terms the values of $\alpha$ tend to be larger.

 In conclusion, we have carried out a tDMRG study of the MIT in an interacting 1D Fermi gas in a quasiperiodic potential. 
The main results of the paper are: a) the dynamics around the MIT is well described by a process of super-diffusion, b) the MIT is not universal -- critical exponents depend on the interaction strength and slowly approach mean-field predictions for sufficiently strong interactions --, c) based on scaling arguments \cite{gang4} we propose that for strong interactions the dynamics tends to ballistic and the localization length $\xi$ diverges  at the MIT as $\xi \propto |\lambda -\lambda_\mathrm{c}|^{-\nu}$ with $\nu \approx 1/2$, d) our results can be tested experimentally in cold atom settings.

 A.M.G. acknowledges support from PTDC /FIS/111348/2009, a Marie Curie International Reintegration Grant
No. PIRG07-GA-2010-26817 and an EPSRC grant No. EP/I004637/1.
The work of M.T. was partially supported by the Grant-in-Aid for the
 Global COE Program ``The Next Generation of Physics, Spun from
 Universality and Emergence'' from MEXT of Japan.

\end{document}